\documentclass[referee]{raa}            

\usepackage{graphicx,times}             
\usepackage{natbib}
\usepackage{amssymb,amsmath}
\bibpunct{(}{)}{;}{a}{}{,}

\usepackage[a4paper=true,dvipdfm=true,pagebackref=true]{hyperref}
\hypersetup{colorlinks = true, linkcolor = green, anchorcolor = red, citecolor = blue, filecolor = red, pagecolor = red, urlcolor = red}

\begin{document}
\title{The r-mode instability windows of strange stars
}

   \volnopage{Vol.19 (2019) No.0, 000--000}      
   \setcounter{page}{1}          

   \author{Yu-Bin Wang
      \inst{1,2}
   \and Xia Zhou
      \inst{2,3}
   \and Na Wang
      \inst{2,3}
    \and Xiong-Wei Liu
    \inst{1}
   }

   \institute{Physics and Space Science College, China West Normal University, Nanchong 637002, China; \\
        \and
             Xinjiang Astronomical Observatory, Chinese Academy of Sciences, Urumqi 830011, China {\it zhouxia@xao.ac.cn}\\
        \and
            Key Laboratory of Radio Astronomy, Chinese Academy of Sciences, Beijing 100049, China\\
\vs\no
   {\small Received~~2018 August 14; accepted~~2018~~October 10}}

\abstract{With the constraint from gravitational wave emission of a binary merger system (GW170817) and two-solar-mass pulsar observations, we investigate the r-mode instability windows of strange stars with unpaired and color-flavor-locked phase strange quark matter. Shear viscosities due to surface rubbing and electron-electron scattering are taken into account in this work. The results show that the effects of the equation of state of unpaired strange quark matter are only dominant at low temperature, but do not have significant effects on strange stars in the color-flavor-locked phase. A color-flavor-locked phase strange star, which is surrounded by an insulating nuclear crust, seems to be consistent with observational data of young pulsars. We find that an additional enhanced dissipation mechanisms might exist in SAX J1808.4-3658. Fast spinning young pulsar PSR J0537-6910 is a primary source for detecting gravitational waves from a rotating strange star, and young pulsars might be strange stars with color-flavor-locked phase strange quark matter.
}

   \authorrunning{Y.-B. Wang, X. Zhou, N. Wang \& X. -W. Liu}            
   \titlerunning{R-mode instability windows of strange star }  

   \maketitle

%
%
\section{Introduction}\label{Sect.1}

The r-modes are non-radial pulsations of compact stars that are primarily driven by
Coriolis forces and are coupled to gravitational radiation (\citealt{Chandrasekhar1970,Friedman1978,Andersson1998,Friedman1998,Owen1998,Andersson2002,Alford2014a}). The r-mode
instability only happens in a range of spin frequencies and temperatures, which is determined by a competition between the effects of gravitational radiation and viscous dissipation damping on modes (\citealt{Lindblom1998}). Therefore, r-mode
instability is an important primary physical mechanism that can prevent pulsars from spinning
up to their Kepler frequency, which also affects the spinning evolutions
of pulsars (\citealt{Madsen1998,Andersson1999,Wang&Dai2017}). Meanwhile, gravitational waves emitted during the instability process could
be detected (\citealt{Owen1998,Andersson2002,Mahmoodifar2013,Alford2014a}).

Such r-modes are damped by viscous dissipation mechanisms (\citealt{Lindblom1998}), and therefore are connected with microscopic properties
of matter inside the stars, which depend on the low energy degrees of freedom and the equation of state (EOS), thus affecting the macroscopic and observable properties of the star. Many studies have tried to constrain the physics behind r-mode instability in compact stars, especially the EOS of cold dense matter, by comparing the r-mode instability windows with the spin frequency and surface temperature of these systems (\citealt{Becker2009,Haskell2012,Mahmoodifar2013,Gusakov2014,Pi2015,Kantor2016,Chugunov2017}).

The LIGO/Virgo detection of gravitational waves from a binary merger system, GW170817, has put a clean and strong constraint on the tidal deformability of the merging objects (\citealt{Abbott2017}).
Combined with two-solar-mass pulsar observations and tidal deformability constraints, \cite{Zhou2018} constrains the parameters of the quark star EOS. In this paper, we will investigate the r-mode instability window of strange stars with unpaired and color-flavor-locked (CFL) phase strange quark matter (SQM). We will utilize a realistic EOS and the new limit on EOS parameters to discuss its effect on the r-mode instability windows of strange star. In the following, we adopt the abbreviations USS for strange star with unpaired SQM, and CSS for strange star with CFL phase SQM.

Most of the uncertainty related to r-mode instability has to do with its damping mechanisms. In a minimum physics model that assumes dissipation only due to standard shear and bulk viscosity, the r-mode instability should be operating in a large portion of the low mass X-ray
binary (LMXB) population and in some young sources (\citealt{Kokkotas2016}). Some works showed that the minimum frequency of the instability region of USS $(\nu_{\rm min}\approx 200 \,\rm Hz)$ is higher than the rotational frequencies of all known young pulsars, and USS can be consistent with both the observed radio and X-ray data (\citealt{Alford2014b}), except for some special sources. CSS without a crust would have a critical frequency at which the r-mode instability sets in measured in Hz or fractions of Hz, in gross disagreement with the data on spin frequencies of both X-ray and radio pulsars (\citealt{Alford2008}). Strange stars can support a thin nuclear crust (\citealt{Glendenning1992}), which only leads to minor changes in the maximum mass compared with bare strange stars (\citealt{Zdunik2002}). The crust on a USS does not have significant effects on the dissipation of r-modes (\citealt{Andersson2002}). As other sources of viscosity are exponentially suppressed in the case of the CFL phase, some viscosity results from the crust might become dominant in the case of the CFL phase, such as shear due to electron-electron scattering or by surface rubbing at low temperature (\citealt{Madsen2000}). These viscosities may have significant effects on the r-mode instability windows of a CSS, leading its behavior to agree with observational data of young radio pulsars. Thus, we will investigate the effect of shear viscosity due to electron-electron scattering or by surface rubbing, by comparing the r-mode instability windows of CSS with observational data of young pulsars, and the USS without a crust will be reviewed by comparing with the observational data of LMXBs.

The paper is organized as follows: In Section \ref{Sect.2}, we briefly review the EOS of unpaired and CFL phase SQM, and give reasonable parameters which are constrained by observations. The main results are presented in Section \ref{Sect.3}, where r-mode instability windows of USS and CSS are provided with different parameters for the EOS. All of the windows are confronted with the tidal deformability measurement from GW170817 and two-solar-mass pulsar observations for systematic constraints on EOS parameters. Moreover, we compare the theoretical r-mode instability windows with the spin frequency and temperature of compact stars in LMXBs and young pulsars. Finally, conclusions and discussion are given in Sec.\ref{Sect.4}

\section{EOS models of strange stars}
\label{Sect.2}

We use the simple but widely used MIT model to describe the unpaired SQM as a mixture of quarks $\rm(u, d, s)$ and electrons (e). The grand canonical potential per unit volume can be written as (\citealt{Weissenborn2011,Bhattacharyya2016})
\begin{equation}
\Omega_{\rm{QM}} = \sum_{i=u,d,s,e} \Omega_{i}^{0}  + \frac{3 \mu^4}{4 \pi^2 } (1-a_{4}) +B_{\rm{eff}},
\end{equation}
where $\Omega_{i}^{0}$ is the grand canonical potential of quarks (u,d,s) and electrons (e) which are regarded as an ideal relativistic Fermi gas. The second term is characterized by the perturbative quantum chromodynamics (QCD) corrections of gluon mediated quark interactions to $O(\alpha_{s}^{2})$ \citep{Fraga2001,Bhattacharyya2016} and the parameter $a_{\rm 4}=1-2 \alpha_{\rm s} /\pi$ represents the degree of quark interaction correction in perturbative QCD (\citealt{Alford2008}).
$\mu=(\mu_{\rm u}+\mu_{\rm d}+\mu_{\rm s})/3.0$ is the baryon chemical potential with the total baryon number density $n_{\rm B}=(n_{\rm u}+n_{\rm d}+n_{\rm s})/3$.
$B_{\rm{eff}}$ is the effective bag constant which includes non-perturbative QCD effects in a phenomenological way.

In the calculation we take $\rm m_u=m_d=0$ and $\rm m_e=0$ (\citealt{Weissenborn2011,Bhattacharyya2016,Li2017,Zhou2018}). The current mass of strange quark $m_{\rm{s}}$ has been constrained well with a recent result of $96 ^{+8}_{-4}\,\rm{MeV}$ \citep{Aaij2016}.
The effect of introducing a finite strange quark mass has been discussed in \cite{Zhou2018} and the results show that a larger $m_{\rm{s}}$ will soften the EOS and $\Lambda(1.4)$ only weakly depends on $m_{\rm{s}}$. In the following, we set $m_{\rm{s}}$ as $100\,\rm{MeV}$.

At large densities, such that up, down and strange quarks bind together, the condensation term, contained in $3 \Delta^2 \mu^2 /\pi^2$, emerges to decrease the free energy of quarks, which are assumed to undergo pairing and form the so-called CFL phase (\citealt{Rajagopal2001}).
As a result, $\Omega_{\rm{CFL}}$ can be obtained from (\citealt{Alford2001,Weissenborn2011})
\begin{equation}
\Omega_{\rm{CFL}} = \Omega_{\rm{QM}}-\frac{3}{\pi^2} \Delta^2 \mu^2,
\end{equation}
where $\Delta$ is the pairing energy gap for the CFL phase, which lacks an accurate calculation within a typical range ($0 - 150\,\rm{MeV}$).

Different choices of parameters for the effective bag constant ($B_{\rm{eff}}$ ), the perturbative QCD correction parameter ($a_4$) and the pairing energy gap ($\Delta$) will lead to various EOS models. These parameters are all confronted with the tidal deformability ($\Lambda$) measurement from GW170817 and other pulsar observations for systematic constraints on those parameters. As discussed in \cite{Zhou2018}, normal quark matter model parameters are compatible with $B_{\rm{eff}}^{1/4}\in (134.1, 141.4)\,\rm{MeV}$ and $a_4\in(0.56,0.91)$ which are derived by  considering the GW170817 constraint on $\Lambda(1.4)$\footnote{Since the high-spin case becomes more loosely bound, we are not going to use this constraint on the parameters.} and two-solar-mass constraint on $M_{\rm{TOV}}$ as well as the stability window for quark matter(``two flavor line'' and ``three flavor line''). The possible constraint on the pairing energy gap $\Delta$ might be larger than $50\,\rm{MeV}$ in the case of $a_4= 1$, and no new lower limit is found for the gap parameter $\Delta$ in the case of $a_4=0.61$ (\citealt{Zhou2018}).

\section{THE R-MODE INSTABILITY WINDOWS OF STRANGE STARS}
\label{Sect.3}
The r-mode instability window is defined as
\begin{equation}
\frac{1}{\tau_{\rm{gw}}} + \frac{1}{\tau_{\rm{V}}} = 0,
\end{equation}
and yields a curve that depends on spin frequency and core temperature. The resulting instability parameter space is commonly depicted as a ``window" in the $\nu-T$ plane. Above this curve, the instability condition is satisfied and the star can emit gravitational waves. $\tau_{\rm{gw}}$ is the growth time due to emission of gravitational waves (\citealt{Lindblom1998,Owen1998})
\begin{equation}
    \frac{1}{\tau_{\rm{gw}}} = - \frac{32 \pi G\Omega^{2l+2}}{c^{2l+3}} \frac{(l-1)^{2l}}{[(2l+1)!!]^2} \bigg(\frac{l+2}{l+1}\bigg)^{2l+2} \int_{0}^{R} \rho r^{2l+2} dr,
\end{equation}
where $G$ is the gravitational constant and $\rho$ is the stellar density.
The viscous damping timescale $\tau_{\rm{V}}$ is given by
\begin{equation}
\frac{1}{\tau_{\rm{V}}}=\sum_{\rm i}\frac{1}{\tau_{\rm i}},
\end{equation}
where the summation is over various dissipation channels(denoted with``$i$''). In the most common case,
in which several dissipation mechanisms are operating simultaneously the viscous timescales are combined
according to the parallel resistors rule. For the simplest
strange star models, two kinds of viscosity are normally considered: bulk viscosity $\tau_{\rm{bv}}$ and shear viscosity $\tau_{\rm{sv}}$.

Bulk viscosity is written as (\citealt{Heiselberg1993,Lindblom1999,Lindblom2002,Nayyar2006})
\begin{equation}
\frac{1}{\tau_{\rm{bv}}} = \frac{4\pi}{690} \bigg(\frac{\Omega^2}{\pi G \bar{\rho}}\bigg)^2 R^{2l-2} \bigg[\int_0^R \rho r^{2l+2} dr\bigg]^{-1}  \int_{0}^{R} \zeta \bigg(\frac{r}{R}\bigg)^6 \bigg[1+0.86 \bigg(\frac{r}{R} \bigg)^{2} \bigg] r^{2} dr,
\end{equation}
where $R$ is the stellar radius and $\bar{\rho} =M/(4 \pi R^3 /3)$ is the averaged density of a non-rotating star (\citealt{Madsen1992}). The bulk viscosity of unpaired SQM mainly depends on the rate of non-leptonic weak interaction: $\rm u+d \leftarrow u+s$. In the high temperature limit ($T>10^9\,\rm K$) (\citealt{Madsen2000})
\begin{eqnarray}
\zeta^{\rm{high}} \approx 3.8 \times 10^{28} m_{100}^{4} \rho_{15}^{-1} T_{9}^{-2} \,\rm g\, cm^{-1}\, s^{-1},
\end{eqnarray}
the bulk viscosity in the low-temperature limit($T<10^9\rm\, K$) is (\citealt{Madsen2000})
\begin{equation}
\zeta^{\rm{low}} \approx 3.2 \times 10^{28} m_{100}^{4} \rho_{15} T_{9}^2 (\kappa \Omega)^{-2} \,\rm g\, cm^{-1}\, s^{-1},
\end{equation}
where $m_{100}$ is the strange quark mass ($m_s$) in the unit of $100\,\rm{MeV}$. For the dominant r-mode ( $m = l = 2$), $\kappa=2/3$.

The dissipation timescale of shear viscosity is written as (\citealt{Lindblom1998})
\begin{equation}
\frac{1}{\tau_{\rm sv}} = (l-1)(2l+1) \bigg[\int _{0}^{R}\rho r^{2l+2}dr\bigg]^{-1} \int _{0}^{R}\eta r^{2l}dr,
\end{equation}
The shear viscosity due to quark-quark scattering has been calculated for the case $T\ll \mu$ ($T$ is the temperature and $\mu$ is the quark chemical
potential) which can be presented as (\citealt{Heiselberg1993})
\begin{equation}
\eta \approx 1.7 \times 10^{18} \bigg(\frac{0.1}{\alpha_{\rm s}}\bigg)^{5/3} \rho_{\rm 15}^{14/9} T_{9}^{-5/3} \,\rm g \, cm^{-1} \, s^{-1},
\end{equation}
where $T_{\rm 9}$ is the core temperature in unit of $10^9\,\rm{K}$ and $\rho_{\rm 15}$ are the density of the star in unit of $10^{15}\, \rm{g\,cm^{-3}}$.

Due to pairing in the CFL phase, the shear viscosity induced by quark-quark scattering is suppressed by the factor $e^{\Delta /3k_{\rm{B}} T}$ and bulk viscosity is suppressed by the factor $e^{2\Delta/ k_{\rm{B}} T}$. Indeed, other viscosities
are only dominant when these effects are exponentially suppressed in the CFL phase (\citealt{Madsen2000}).
For CSS, the viscosity at low temperature
is determined by shear due to electron-electron scattering
or by surface rubbing
\begin{equation}
\tau_{\rm sv}^{\rm ee} = 2.95\times 10^7(\mu_e /\mu_q)^{-14/3}T^{5/3}_9 \,\rm s,
\end{equation}
We take the maximized effect of electron shear $\rm \mu_e /\mu_q =0.1$ (\citealt{Madsen2000}). Surface rubbing due to the electron atmosphere being carried along by the r-modes in the quark phase, scattering mainly on photons in the nuclear crust, corresponds to a viscous time scale
\begin{equation}
\tau_{\rm sr} \simeq (1.42\times 10^8)s~T_9 (\nu/1\, \rm{kHz})^{-1/2},
\end{equation}
for a crust with maximal density (\citealt{Madsen2000}).

\begin{figure}
\centering
  \resizebox{\hsize}{!}{\includegraphics[width=70mm]{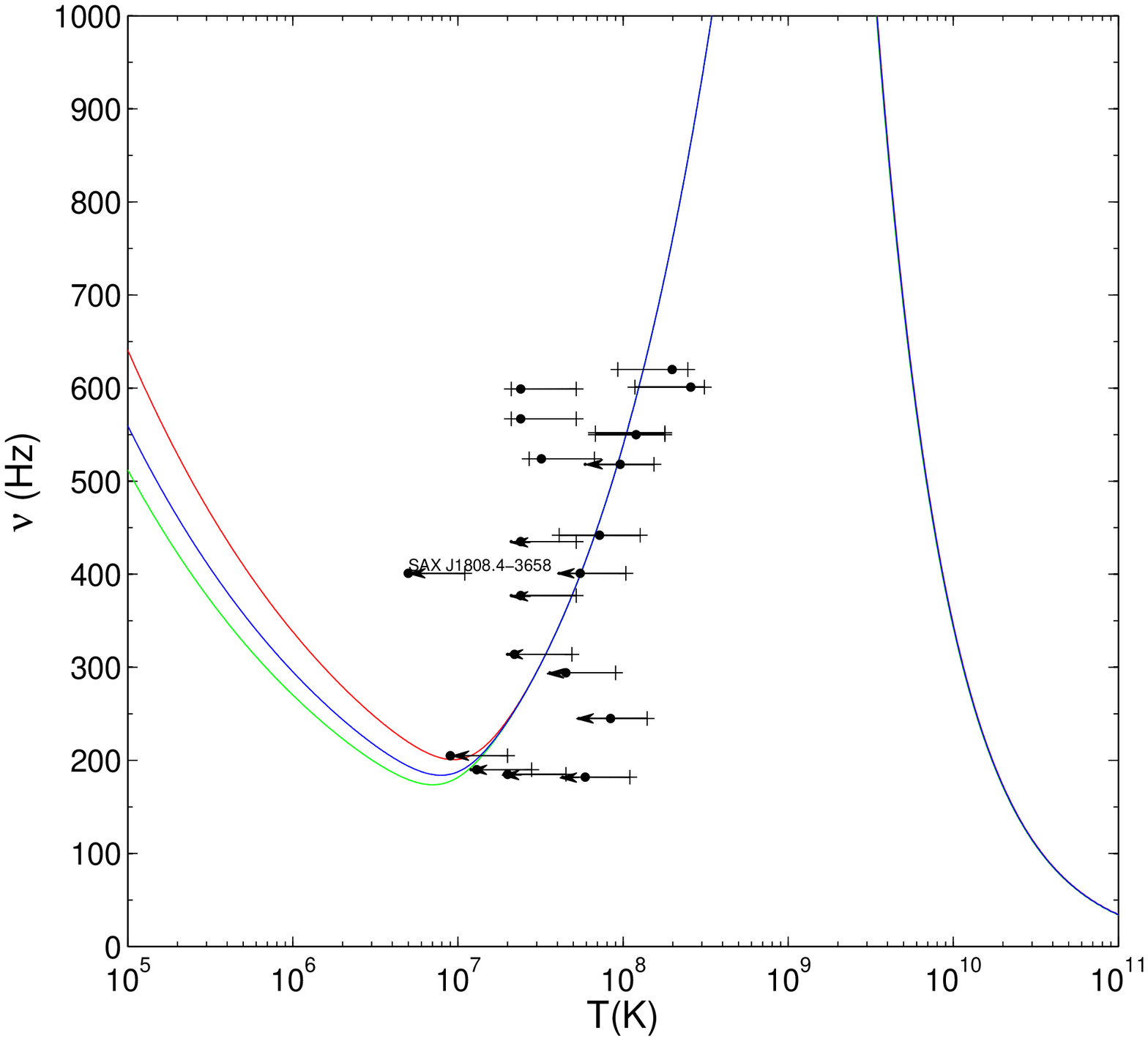} \includegraphics[width=70mm]{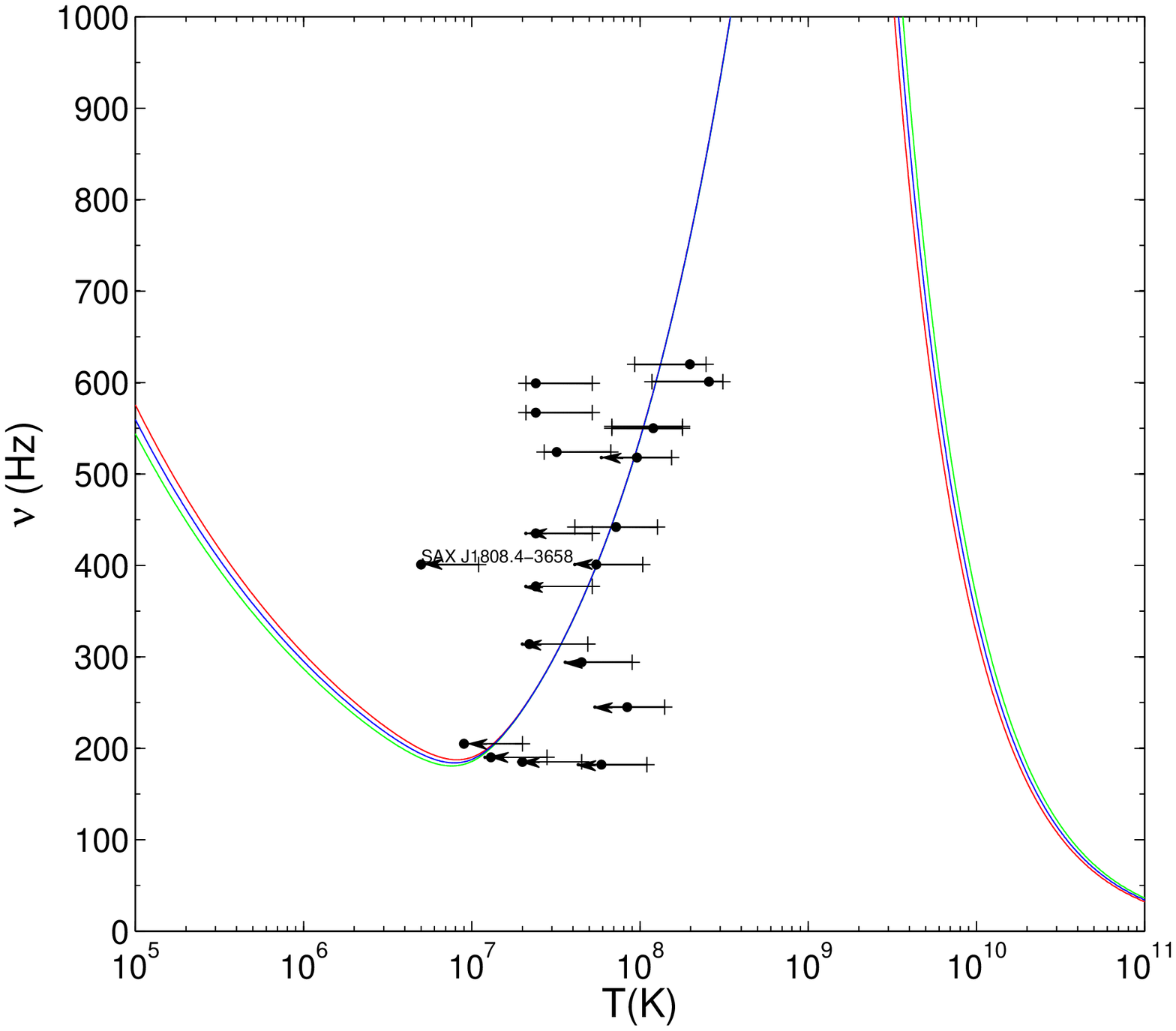}}
  \caption{R-mode instability windows for USS with $M=1.4\,M_{\odot}$. The observed spin frequency and core temperature of LXMBs are also included for a comparison. The left panel is for different $a_4$ with $B_{\rm eff}^{1/4}=138\,{\rm MeV}$ and the right one is for different $B_{\rm eff}^{1/4}$ with $a_4=0.72$. In the left panel, the \emph{green line}, \emph{blue line} and \emph{red line} correspond to $a_4=0.61, 0.72$ and $0.83$ respectively. In the right panel, the \emph{green line}, \emph{blue line} and \emph{red line} signify $B_{\rm eff}^{1/4}=136$, $138$ and $140\,{\rm MeV}$ respectively.}\label{fX1}
\end{figure}

By solving Equation (3) and different damping mechanisms numerically, we can derive the r-mode instability windows for USS and CSS.
Fig.\ref{fX1} shows the r-mode instability windows for USS with $M=1.4\,M_{\odot}$. The instability windows are calculated with different parameters of EOS, which are constrained by two-solar-mass pulsar observations and tidal deformability of the merging binary. Also indicated are the spin frequency and core temperature of compact stars in LMXBs, which have been given in \citealt{Gusakov2014}, and temperature for HETE J1900.1-2455 ($5\times 10^5\, \rm K$) was updated in \citealt{Degenaar2017}. The left panel is for $a_4=0.61, 0.72$ and $0.83$, and $B_{\rm eff}^{1/4}=138\,\rm{MeV}$. The right one is for $B_{\rm eff}^{1/4}=136\,\rm{MeV},138\,\rm{MeV}$ and $140\,\rm{MeV}$, and $a_4=0.72$. We find that $a_4$ and $B_{\rm eff}^{1/4}$ do not have significant effects on the r-mode instability window at high temperature (approximately higher than $10^7 \,\rm K$); r-mode instability windows at low temperature are shrinking while $a_4$ or $B_{\rm eff}^{1/4}$ are increasing. A smaller allowed $a_{\rm 4}$ will lead to a larger r-mode instability window.

\begin{figure}
\centering
  \includegraphics[width=70mm]{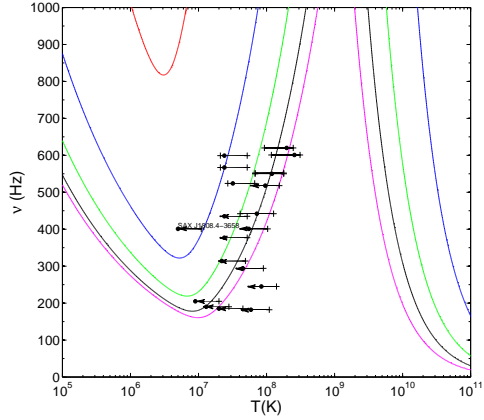}\\
  \caption{Similar to Fig.1,but for USS with $M=0.1\,M_{\odot}$(\emph{red line}), $M=0.5\,M_{\odot}$(\emph{blue line}),$M=1.0\,M_{\odot}$(\emph{green lien}), $M=1.5\,M_{\odot}$(\emph{black line}) and $M=2.0\,M_{\odot}$(\emph{mauve line}). The corresponding parameters for the EOSs are set as $B_{\rm eff}^{1/4}=138\,\rm{MeV}$ and $a_4=0.72$.} \label{fX2}
\end{figure}

Since some pulsar like objects might be low mass strange stars (\citealt{Zhou2017}), we present the r-mode instability windows for USS with different masses in Fig.\ref{fX2}, especially for mass $<1.0\, M_{\odot}$. The windows become larger while the star mass increases. Many sources are around right boundaries of instability windows and SAX J1808.4-3658 is in the instability region even though the star mass is $0.5\,M_{\odot}$. The sources are out of the instability region when the mass is $0.1\, M_{\odot}$.

%

\begin{figure}
\centering
  \resizebox{\hsize}{!}{\includegraphics[width=70mm]{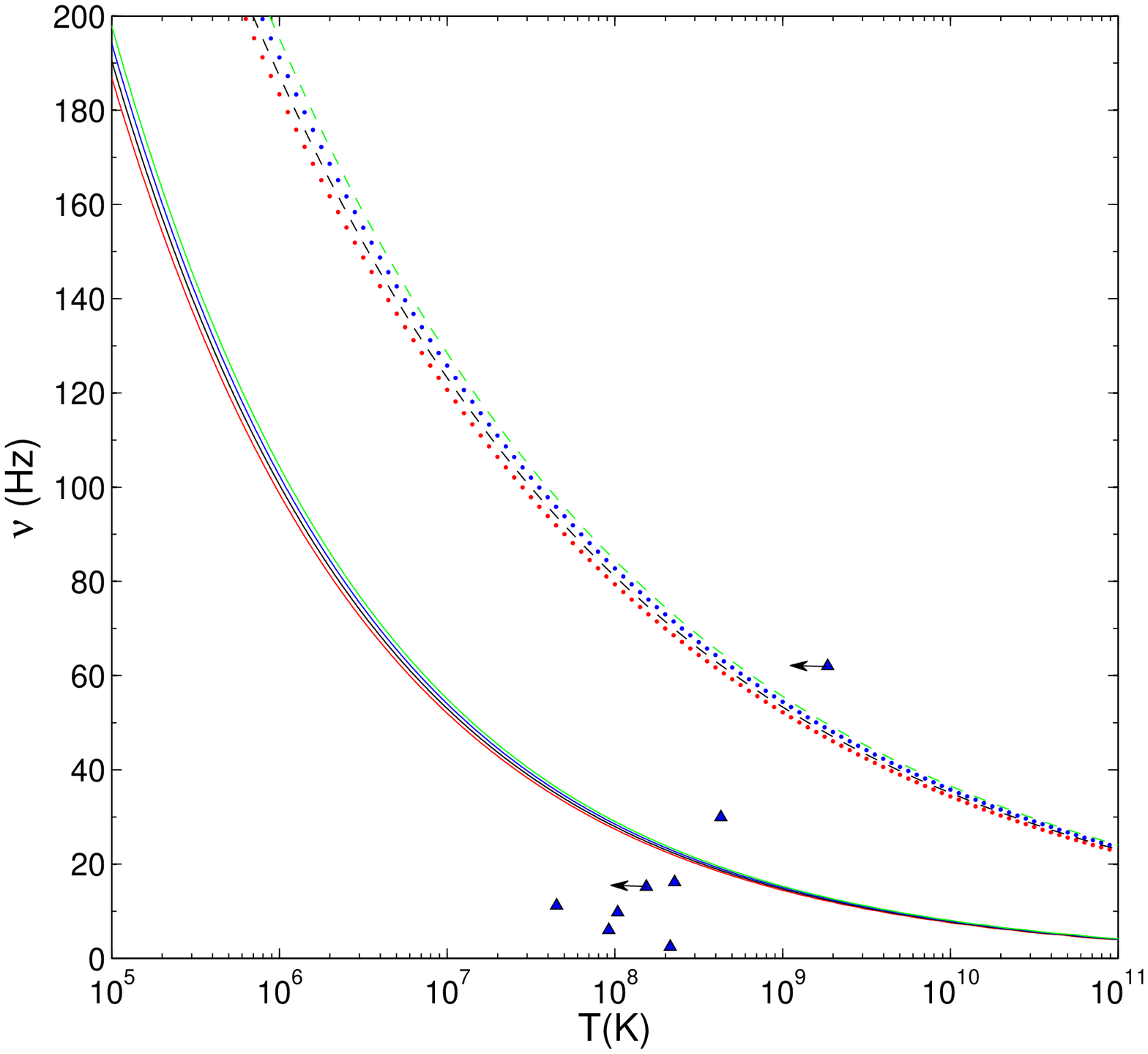}\includegraphics[width=70mm]{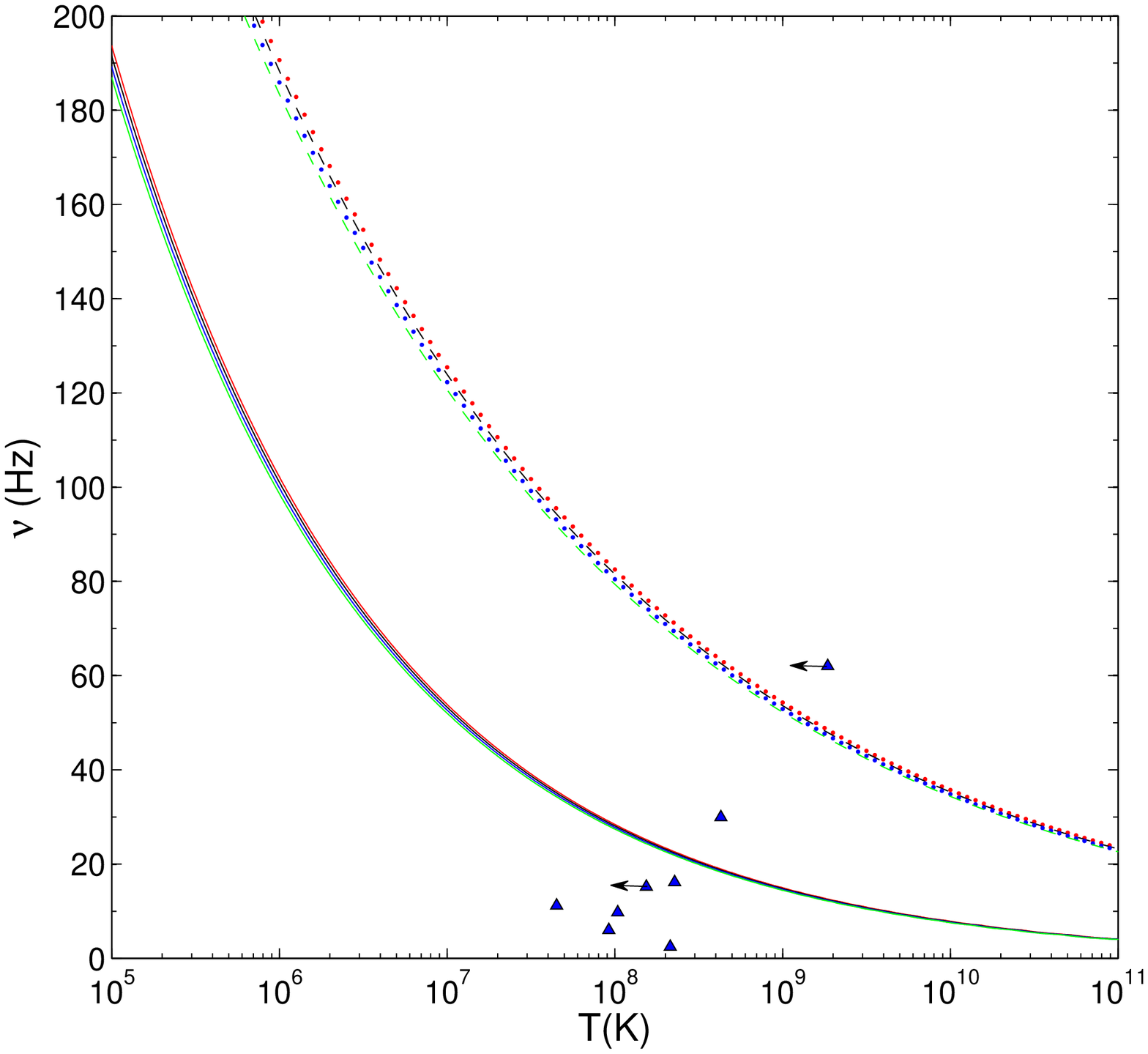}}
  \caption{R-mode instability windows for CSS with $a_4=1$ and $M=1.4\,M_{\odot}$. The observed spin frequency and core temperature of young pulsars tabulated in Table\ref{tb1} are also included for comparison(\emph{blue triangles}). The left panel is for different $B^{1/4}_{\rm{eff}}$ with $\Delta=100\,\rm{MeV}$, and the \emph{red, black, blue} and \emph{green lines} correspond to $B_{\rm eff}^{1/4}=146, 148.5, 151, 153.5\,\rm{MeV}$ respectively. The right panel is for different $\Delta$ with $B_{\rm eff}^{1/4}=146\,\rm{MeV}$, and the \emph{red, black, blue} and \emph{green lines} represent $\Delta=70, 80, 90, 100\,\rm{MeV}$ respectively. \emph{Dotted curves} or \emph{dashed curves} mean shear due to surface rubbing, and \emph{solid curves} signify shear due to electron-electron scattering}\label{fY11}
\end{figure}

\begin{figure}
\centering
  \resizebox{\hsize}{!}{\includegraphics[width=70mm]{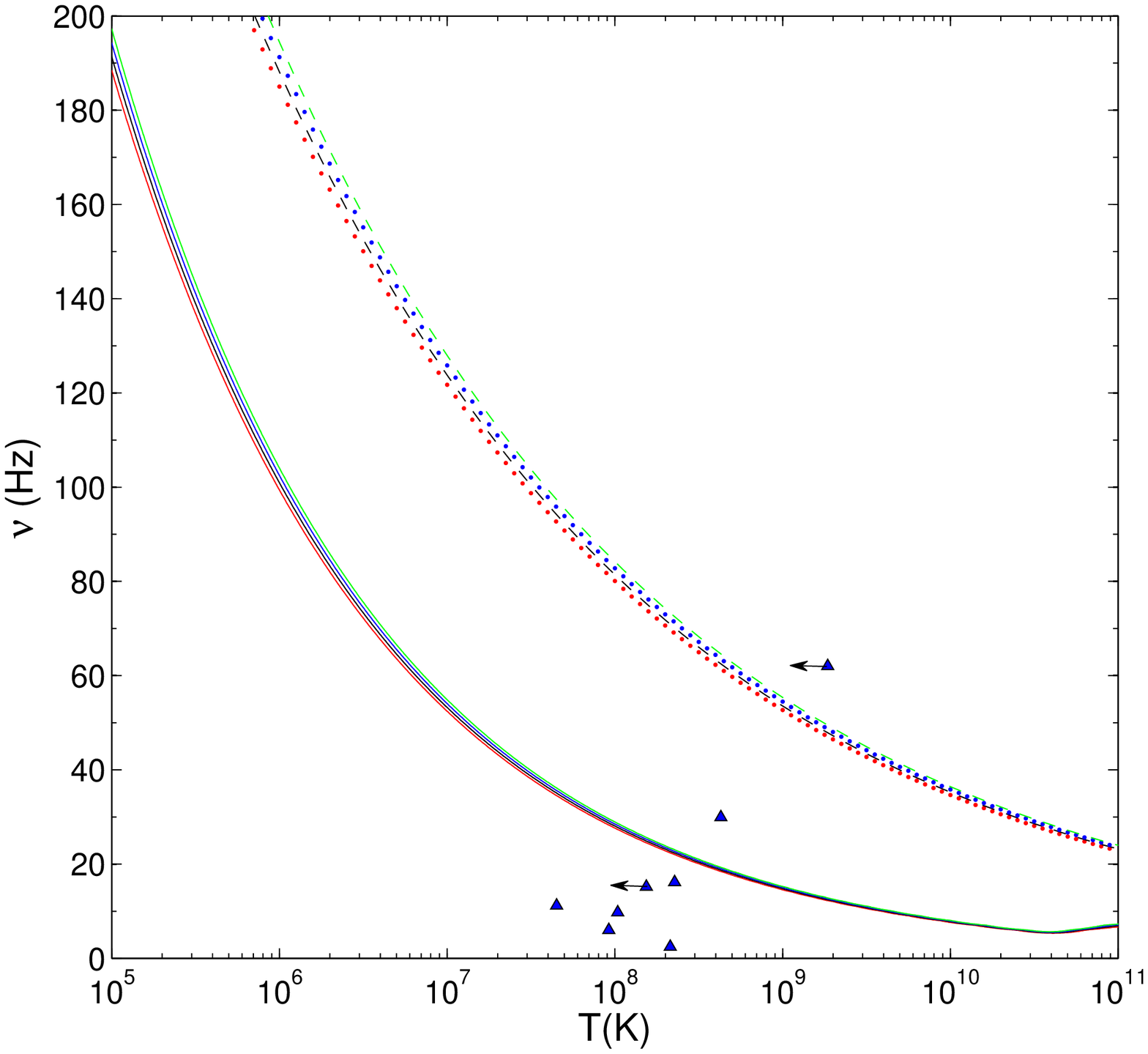}\includegraphics[width=70mm]{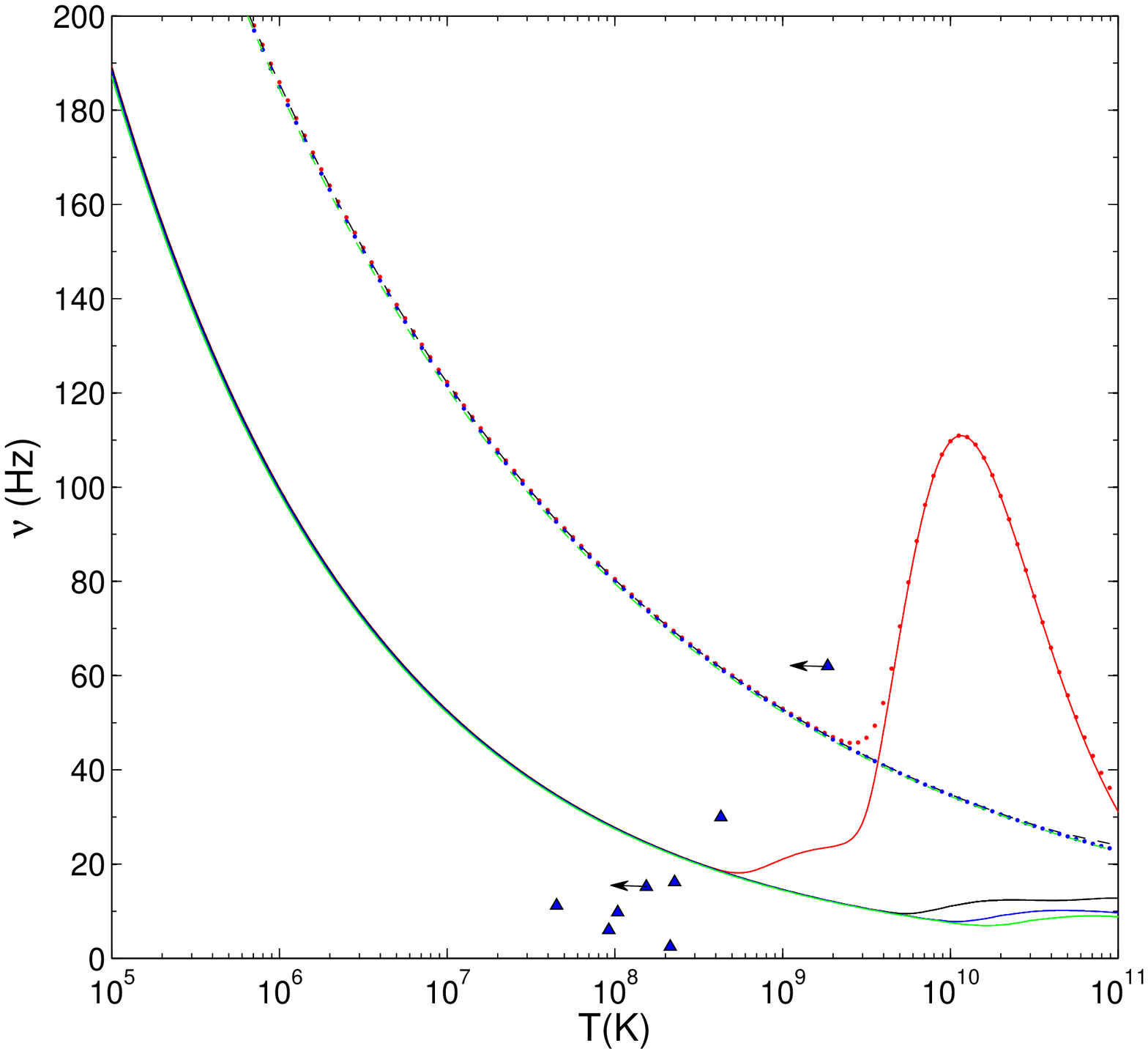}}
  \caption{R-mode instability windows for CSS with $a_4 = 0.61$ and $M = 1.4\,M_{\odot}$. The observed spin frequency and core temperature of young pulsars tabulated in Table\ref{tb1} are also included for comparison(\emph{blue triangles}). \emph{Dotted curves} or \emph{dashed curves} correspond to shear due to surface rubbing and \emph{solid curves} indicate shear due to electron-electron scattering. The left panel is for different $B_{\rm eff}^{1/4}$ with $\Delta=75\,{\rm MeV}$, and the \emph{red , black, blue} and \emph{green lines} represent $B_{\rm eff}^{1/4}=144, 146, 148, 150\,{\rm MeV}$ respectively. The right panel is for different gaps with $B_{\rm eff}^{1/4}=136.5\,{\rm MeV}$, and the \emph{red, black, blue} and \emph{green lines} correspond to $\Delta= 1, 10, 20, 30\,\rm{MeV}$ respectively.}\label{fY21}
\end{figure}

The r-mode instability windows for CSS with different gaps and effective bag constants are displayed in Fig.\ref{fY11} and Fig.\ref{fY21}. The spin frequency and core temperature of young pulsars are listed in Table\ref{tb1}, and their core temperature has been inferred by assuming an outer envelope completely composed of a crust of iron elements (\citealt{Becker2009}). We consider the shear due to surface rubbing and electron-electron scattering respectively in Fig. \ref{fY11} and \ref{fY21}. It is shown that the suppressed effect of surface rubbing for the instability window is much stronger than that for shear by electron-electron scattering. In addition, the viscous timescale for shear due to surface rubbing is $\tau_{sr} \propto (10^8-10^{10})T_9$ and the viscous timescale for shear due to electron-electron scattering is $\rm\tau_{sv}^{ee} \propto 10^{12}T_{9}^{5/3}$. This implies that the dominant role in suppressing the r-modes is supposed to be the shear due to surface rubbing. In order to show the damping effect from the existence of a crust, we consider the shear due to either surface rubbing
or electron-electron scattering in the following descriptions.

In Fig. \ref{fY11} and \ref{fY21}, most of the young pulsars are located outside the instability region, except for the fast spinning PSR J0537-6910 and the Crab. Different gaps as well as different effective bag constants and $a_4$ lead to a slight influence on the r-mode instability windows of CSS.The windows at high temperature dramatically shrink when the gaps decrease to $\Delta=1\,\rm MeV$, which are shown in the right panel in Fig.\ref{fY21}. No matter if shear due to surface rubbing or electron-electron scattering is considered, PSR J0537-6910 is always in the instability region, but when a star damped by shear due to surface rubbing is examined, Crab is out of the instability windows.

\begin{figure}[h]
\centering
  \resizebox{\hsize}{!}{\includegraphics[width=70mm]{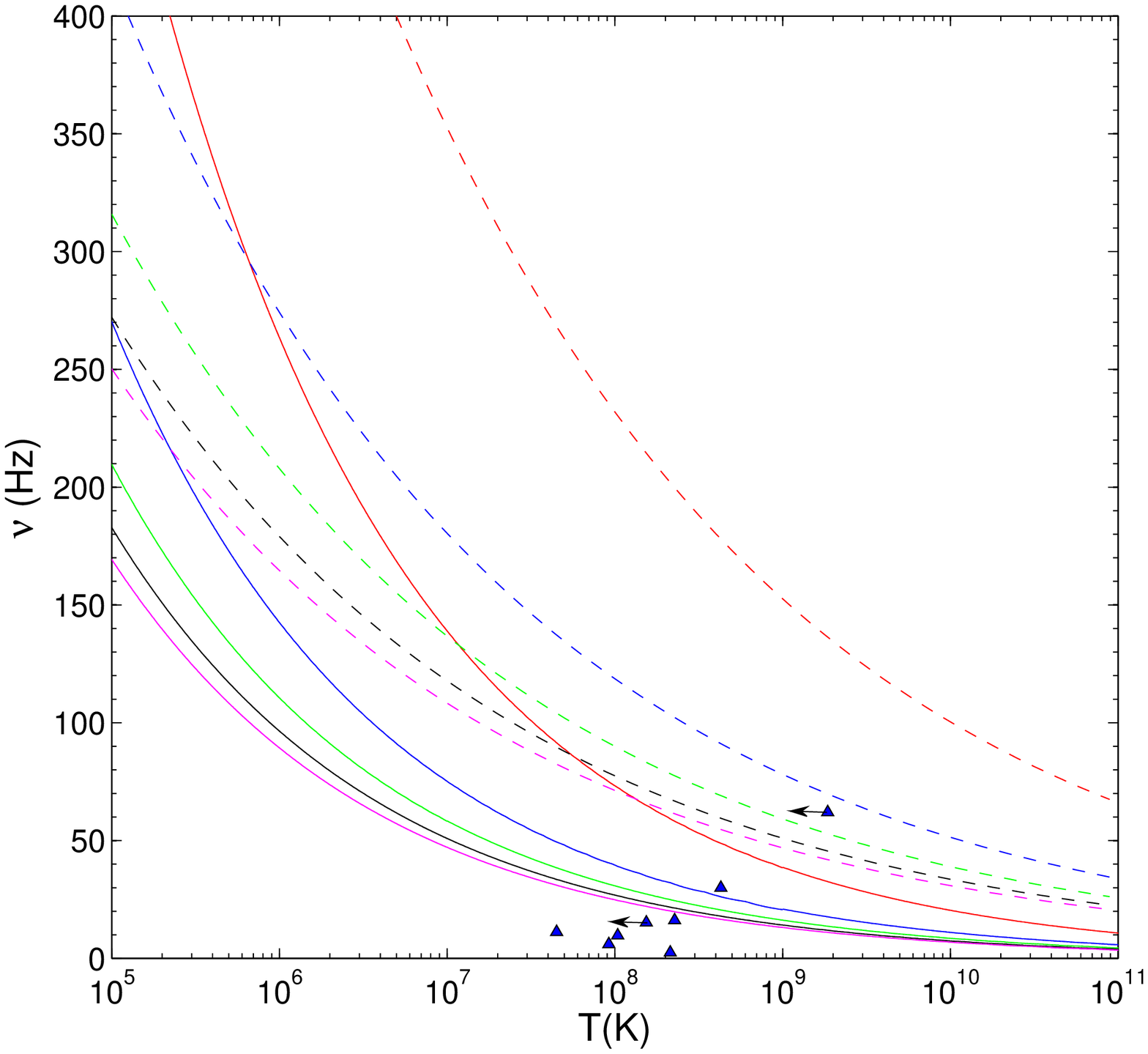}\includegraphics[width=70mm]{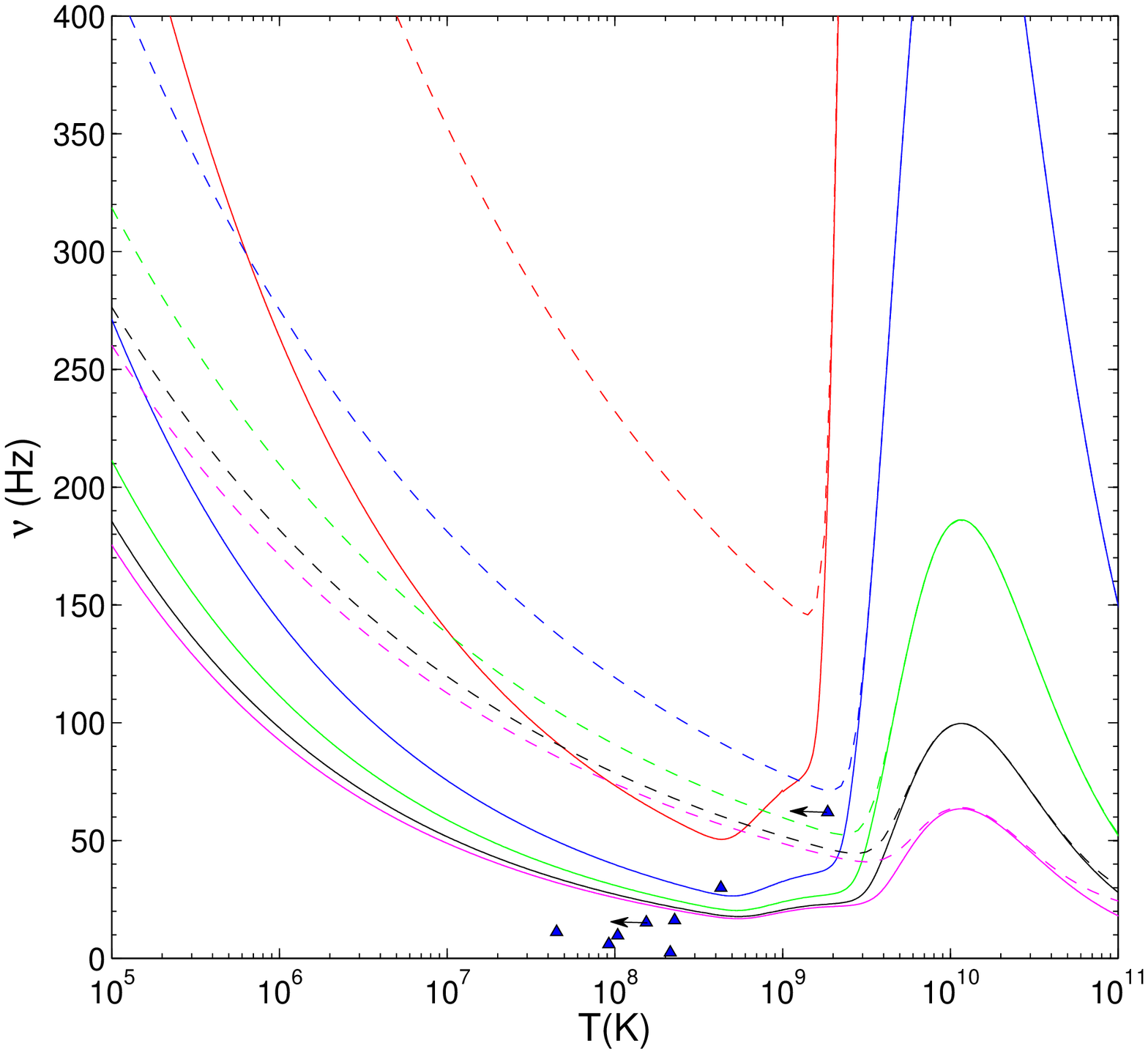}}
  \caption{R-mode instability windows for CSS with different masses. The observed spin frequency and core temperature of young pulsars tabulated in Table\ref{tb1} are also included for comparison(blue triangles). The left panel is for $a_4 = 1, \Delta=100\,{\rm MeV}$ and $B_{\rm eff}^{1/4}=146\,{\rm MeV}$. The right panel is for $a_4 = 0.61$, $\Delta=1\,{\rm MeV}$ and $B_{\rm eff}^{1/4}=136.5\,{\rm MeV}$. The \emph{red, blue, green, black} and \emph{mauve lines} correspond to $M=0.1\,M_{\odot}$, $M=0.5\,M_{\odot}$, $M=1.0\,M_{\odot}$, $M=1.5\,M_{\odot}$, $M=2.0\,M_{\odot}$ respectively. \emph{Dashed} (\emph{solid}) \emph{curves} signify shear due to surface rubbing (electron-electron scattering).} \label{fY31}
\end{figure}

In Fig.\ref{fY31}, the r-mode instability windows for CSS with different star masses($M=0.1, 0.5, 1.0, 1.5, 2.0\,M_{\odot}$) are presented. It is found that the instability windows monotonically increase with the star mass gradually increasing. With small mass($<1.0\,M_{\odot}$), the two fast spinning pulsars might be outside of these windows.

\begin{table*}
  \centering
  \caption{\label{table:noktaatisi} Spin frequency and surface temperature for young pulsars that have measurements (or upper limits) of both. Spin frequency for young pulsars are taken from the ATNF catalog (\citealt{Manchester2005}).}\label{tb1}
  \begin{minipage}{\linewidth}
  \begin{center}
  \begin{tabular}{lcrcr}
  \hline\noalign{\smallskip}
  $\rm Source$ & $\nu$ & $T_{s}^{\rm \infty}$  & $T_{\rm core}(\rm Fe)$& \rm References\\
   $ $         &  (Hz) &$(10^6\, \rm K)$       & $(10^8\, \rm K)$      &  $ $\\
  \hline\noalign{\smallskip}
  PSR B0531+21(Crab)& 29.95 &  1.79 &  4.29  & \citealt{Weisskopf2004}\\
  PSR J0205+6449    & 15.22 &$<1.02$&  1.54  & \citealt{Slane2004}\\
  PSR J1119-6127    & 2.451 &  1.22 &  2.14  & \citealt{Safi-Harb2008}\\
  PSR J1357-6429    & 6.020 &  0.77 &  0.92  & \citealt{Zavlin2007}\\
  PSR B0833-45(Vela)& 11.20 &  0.52 &  0.45  & \citealt{Manzali2007}\\
  PSR B1706-44      & 9.760 &  0.82 &  1.04  & \citealt{McGowan2004}\\
  PSR J0537-6910    & 62.03 &$<4.00$&  18.58 & \citealt{Andersson2018}\\
  PSR J1833-1034    & 16.16 &  1.26 &  2.27  & \citealt{Matheson2010}\\
  \noalign{\smallskip}\hline
  \end{tabular}
  \end{center}
  \end{minipage}
\end{table*}

\section{CONCLUSIONS and Discussion}
\label{Sect.4}

Coupled with a new constraint on EOS, we investigate the r-mode instability windows of USS and CSS. Moreover, these r-mode instability windows are compared with the spin frequency and core temperature of compact star in LMXBs/millisecond pulsars (MSPs) and young pulsars. The effects of EOS for unpaired SQM are only dominant at low temperature ($T\lesssim 10^7 \,\rm K$). For CSS, the damping mechanisms due to the existence of the crust are important for us to consider when drawing any conclusions.

R-mode instability has been identified as a viable and promising target for gravitational wave searches and has been taken into account in recent analysis. LMXBs and young pulsars are possible sources for detecting gravitational waves driven by r-mode (\citealt{Kokkotas2016}). USS seems consistent with LMXB data except for SAX J1808.4-3658.
This source is located in the instability region unless the mass is smaller than $0.5\,M_{\odot}$ and therefore could experience
r-mode-driven spin-down. However, the measured spin-down rate is consistent with that caused by a canonical LMXB
magnetic field ($\sim 10^8\, \rm G$), hence suggesting that the r-mode instability is not the dominant effect and SAX J1808.4-3658 may be a massive star (\citealt{Salmi2018}). This apparent tension between the minimum damping r-mode model of USS and the spin-temperature-mass data of SAX J1808.4-3658 may be in conflict with each other. That means additional damping mechanisms might work in this system, which could modify the instability window.

The shear viscosity due to quark scattering and bulk viscosity are suppressed in the CFL phase, which lead to disagreement with the observational data of LMXBs (\citealt{Alford2008}). Actually, a CSS may have a tiny nuclear crust (\citealt{Madsen2000}). As shown in this work, EOSs do not have a significant effect on the instability windows of CSS. The most effective dissipation mechanisms are the shear due to surface rubbing or electron-electron scattering. PSR J0537-6910 is always located in the instability window no matter if shear due to surface rubbing or electron-electron scattering is considered. From this point of view, the fast spinning young pulsar PSR J0537-6910 is extremely promising source for detecting gravitational waves from a rotating strange star. In addition to this, we could learn something about the structure and mass of PSR J0537-6910 or Crab by comparing the r-mode instability with observational data.

Moreover, a radio observation recently demonstrated that PSR J0537-6910 appears to be close to the LIGO sensitivity for mass $M=1.4\,M_{\odot}$ and radius roughly taken as $10-14\,\rm km$, and its gravitational wave radiation can possibly be detected by advanced LIGO(aLIGO) (\citealt{Andersson2018}). The results above suggest that the case for USS has difficulty in agreeing with the radio observation (\citealt{Andersson2018}). In addition, if the mass of PSR J2215+5135 is confirmed ($M= 2.27_{-0.15}^{+0.17}\,M_{\odot}$), USS would be rule out since the tidal deformability measurement of GW170817 implies $M\leq2.18M_{\odot}$ (\citealt{Linares2018,Zhou2018}). If these case turn out to be ture, young pulsars might be CFL phase strange stars.

\begin{acknowledgements}
We thank the referee for valuable suggestions that have allowed us to improve our manuscript significantly. The work was supported by the National Natural Science Foundation of China Nos. 11373006, 11873040, U1838108, U1531137.
\end{acknowledgements}

\label{lastpage}


\begin{thebibliography}{}

\bibitem[Aaij et al.(2016)]{Aaij2016} Aaij, R., Adeva, B., Adinolfi, M., et al.\ 2016, Chinese Physics C, 40, 011001 %

\bibitem[Abbott et al.(2017)]{Abbott2017} Abbott, B.~P., Abbott, R., Abbott, T.~D., et al.\ 2017, Physical Review Letters, 119, 161101

\bibitem[Alford et al.(2001)]{Alford2001} Alford, M., Rajagopal, K., Reddy, S., \& Wilczek, F.\ 2001, Phys. Rev. D, 64, 074017

\bibitem[Alford et al.(2008)]{Alford2008} Alford, M.~G., Schmitt, A., Rajagopal, K., \& Sch{\"a}fer, T.\ 2008, Reviews of Modern Physics, 80, 1455

\bibitem[Alford \& Schwenzer(2014a)]{Alford2014a} Alford, M.~G., \& Schwenzer, K.\ 2014a, ApJ, 781, 26

\bibitem[Alford \& Schwenzer(2014b)]{Alford2014b} Alford, M.~G., \& Schwenzer, K.\ 2014b, Physical Review Letters, 113, 251102

\bibitem[Andersson(1998)]{Andersson1998} Andersson, N.\ 1998, ApJ, 502, 708 %

\bibitem[Andersson et al.(1999)]{Andersson1999} Andersson, N., Kokkotas, K.~D., \& Stergioulas, N.\ 1999, ApJ, 516, 307

\bibitem[Andersson et al(2002)]{Andersson2002} Andersson, N., Jones, D.~I., \& Kokkotas, K.~D.\ 2002, MNRAS, 337, 1224 %

\bibitem[Andersson et al.(2018)]{Andersson2018} Andersson, N., Antonopoulou, D., Espinoza, C.~M., Haskell, B., \& Ho, W.~C.~G.\ 2018, ApJ, 864, 137

\bibitem[Becker(2009)]{Becker2009} Becker W., 2009, Neutron Stars and Pulsars. (Berlin: Springer) %

\bibitem[Bhattacharyya et al.(2016)]{Bhattacharyya2016} Bhattacharyya, S., Bombaci, I., Logoteta, D., \& Thampan, A.~V.\ 2016, MNRAS, 457, 3101

\bibitem[Chandrasekhar(1970)]{Chandrasekhar1970} Chandrasekhar, S.\ 1970, Physical Review Letters, 24, 611

\bibitem[Chugunov et al.(2017)]{Chugunov2017} Chugunov, A.~I., Gusakov, M.~E., \& Kantor, E.~M.\ 2017, MNRAS, 468, 291

\bibitem[Degenaar et al.(2017)]{Degenaar2017} Degenaar, N., Ootes, L.~S., Reynolds, M.~T., Wijnands, R., \& Page, D.\ 2017, MNRAS, 465, L10

\bibitem[Fraga et al.(2001)]{Fraga2001} Fraga, E.~S., Pisarski, R.~D., \& Schaffner-Bielich, J.\ 2001, Phys. Rev. D, 63, 121702 %

\bibitem[Friedman \& Schutz(1978)]{Friedman1978} Friedman, J.~L., \& Schutz, B.~F.\ 1978, ApJ, 222, 281 %

\bibitem[Friedman \& Morsink(1998)]{Friedman1998} Friedman, J.~L., \& Morsink, S.~M.\ 1998, ApJ, 502, 714  %

\bibitem[Glendenning \& Weber (1992)]{Glendenning1992} Glendenning, N.~K., \& Weber, F.\ 1992, ApJ, 400, 647 %

\bibitem[Gusakov et al.(2014)]{Gusakov2014} Gusakov, M.~E., Chugunov, A.~I., \& Kantor, E.~M.\ 2014, Phys. Rev. D, 90, 063001

\bibitem[Haskell et al.(2012)]{Haskell2012} Haskell, B., Degenaar, N., \& Ho, W.~C.~G.\ 2012, MNRAS, 424, 93 %

\bibitem[Heiselberg(1993)]{Heiselberg1993} Heiselberg, H., \& Pethick, C.~J.\ 1993, Phys. Rev. D, 48, 2916 %

\bibitem[Hessels et al.(2006)]{Hessels2006} Hessels, J.~W.~T., Ransom, S.~M., Stairs, I.~H., et al.\ 2006, Science, 311, 1901

\bibitem[Kantor et al.(2016)]{Kantor2016} Kantor, E.~M., Gusakov, M.~E., \& Chugunov, A.~I.\ 2016, MNRAS, 455, 739 %

\bibitem[Kokkotas \& Schwenzer (2016)]{Kokkotas2016} Kokkotas, K.~D., \& Schwenzer, K.\ 2016, European Physical Journal A, 52, 38 %

\bibitem[Li et al.(2017)]{Li2017} Li, A., Zhu, Z.-Y., \& Zhou, X.\ 2017, ApJ, 844, 41

\bibitem[Linares et al.(2018)]{Linares2018} Linares, M., Shahbaz, T., \& Casares, J.\ 2018, ApJ, 859, 54

\bibitem[Lindblom et al.(1998)]{Lindblom1998} Lindblom, L., Owen, B.~J., \& Morsink, S.~M.\ 1998, Physical Review Letters, 80, 4843 %

\bibitem[Lindblom et al.(1999)]{Lindblom1999} Lindblom, L., Mendell, G., \& Owen, B.~J.\ 1999, Phys. Rev. D, 60, 064006 %

\bibitem[Lindblom \& Owen(2002)]{Lindblom2002} Lindblom, L., \& Owen, B.~J.\ 2002, Phys. Rev. D, 65, 063006

\bibitem[Madsen(1992)]{Madsen1992} Madsen, J.\ 1992, Phys. Rev. D, 46, 3290

\bibitem[Madsen(1998)]{Madsen1998} Madsen, J.\ 1998, Physical Review Letters,  81, 3311 %

\bibitem[Madsen(2000)]{Madsen2000} Madsen, J.\ 2000, Physical Review Letters, 85, 10 %

\bibitem[Mahmoodifar \& Strohmayer(2013)]{Mahmoodifar2013} Mahmoodifar, S., \& Strohmayer, T.\ 2013, ApJ, 773, 140

\bibitem[Manchester et al.(2005)]{Manchester2005} Manchester, R.~N., Hobbs, G.~B., Teoh, A., \& Hobbs, M.\ 2005, AJ, 129, 1993 %

\bibitem[Manzali et al.(2007)]{Manzali2007} Manzali, A., De Luca, A., \& Caraveo, P.~A.\ 2007, ApJ, 669, 570

\bibitem[Matheson \& Safi-Harb(2010)]{Matheson2010} Matheson, H., \& Safi-Harb, S.\ 2010, ApJ, 724, 572

\bibitem[McGowan et al.(2004)]{McGowan2004} McGowan, K.~E., Zane, S., Cropper, M., et al.\ 2004, ApJ, 600, 343

\bibitem[Nayyar \& Owen(2006)]{Nayyar2006} Nayyar, M., \& Owen, B.~J.\ 2006, Phys. Rev. D, 73, 084001 %

\bibitem[Owen et al.(1998)]{Owen1998} Owen, B.~J., Lindblom, L., Cutler, C., et al.\ 1998, Phys. Rev. D, 58, 084020 %

\bibitem[Pi et al.(2015)]{Pi2015} Pi, C.-M., Yang, S.-H., \& Zheng, X.-P.\ 2015, RAA, 15, 871

\bibitem[Rajagopal \& Wilczek(2001)]{Rajagopal2001} Rajagopal, K., \& Wilczek, F.\ 2001, Physical Review Letters, 86, 3492 %

\bibitem[Safi-Harb \& Kumar(2008)]{Safi-Harb2008} Safi-Harb, S., \& Kumar, H.~S.\ 2008, ApJ, 684, 532

\bibitem[Salmi et al.(2018)]{Salmi2018} Salmi, T., N{\"a}ttil{\"a}, J., \& Poutanen, J.\ 2018, preprint (arXiv:1805.01149) %

\bibitem[Slane et al.(2004)]{Slane2004} Slane, P., Helfand, D.~J., van der Swaluw, E., \& Murray, S.~S.\ 2004, ApJ, 616, 403

\bibitem[Wang \& Dai(2017)]{Wang&Dai2017} Wang, J.-S., \& Dai, Z.-G.\ 2017, A\&A, 603, A9

\bibitem[Weissenborn et al.(2011)]{Weissenborn2011} Weissenborn, S., Sagert, I., Pagliara, G., Hempel, M., \& Schaffner-Bielich, J.\ 2011, ApJ, 740, L14

\bibitem[Weisskopf et al.(2004)]{Weisskopf2004} Weisskopf, M.~C., O'Dell, S.~L., Paerels, F., et al.\ 2004, ApJ, 601, 1050

\bibitem[Zavlin(2007)]{Zavlin2007} Zavlin, V.~E.\ 2007, ApJ, 665, L143

\bibitem[Zdunik (2002)]{Zdunik2002} Zdunik, J.~L.\ 2002, A\&A, 394, 641 %

\bibitem[Zhou et al.(2017)]{Zhou2017} Zhou, X., Tong, H., Zhu, C., \& Wang, N.\ 2017, MNRAS, 472, 2403 %

\bibitem[Zhou et al.(2018)]{Zhou2018} Zhou, E.-P., Zhou, X., \& Li, A.\ 2018, Phys. Rev. D, 97, 083015

\end{thebibliography}
\end{document}